\documentclass{article}
\usepackage[utf8]{inputenc}
\usepackage{graphicx,float,amsmath,url,amssymb,amsfonts,bbm,bbding}

\title{Null/No Information Rate (NIR): a statistical test to assess if a classification accuracy is {\itshape significant} for a given problem}
\author{M. Bicego, A. Mensi\\University of Verona (Italy)}

\begin{document}

\maketitle

\begin{abstract}
In many research contexts, especially in the biomedical field, after studying and developing a classification system a natural question arises: ``Is this accuracy {\itshape enough} high?'', or better, ``Can we say, with a statistically significant confidence, that our classification system is able to {\itshape solve} the problem''? To answer to this question we can use the statistical test described in this paper, which is referred in some cases as {\itshape NIR } (No Information Rate or Null Information Rate). 
\end{abstract}

\section{Motivation}
In many research contexts, especially in the biomedical field, we have a classification problem for which we develop a classification system. Then we evaluate the performances of such system by measuring its {\itshape classification accuracy (or error)}, typically estimated with a Cross Validation protocol. In particular we have a dataset, which contains objects for which we know the true category, and we split the dataset in two separated sets: one, called {\itshape training set}, is used to build the classifier and the other, called {\itshape testing set}, is used to test it: we classify the objects in the testing set with the trained classifier and we count the number of times our classifier provides a correct answer, i.e. the answer of the classifier on a given object is identical to its true label.

At this point a simple question may arise: ``Is this accuracy {\itshape enough} high?'', or better, ``Can we say, with a statistically significant confidence, that our classification system is able to {\itshape solve} the problem''? To answer to this question we can use the statistical test described here. In particular we aim at assessing, with a statistical test, if the computed classification accuracy is better than those which would have been obtained when:
\begin{itemize} 
\item we assign every object of the testing set to a class which is {\itshape randomly chosen} 
\item we assign every object of the testing set to the class which is {\itshape more common} in the problem 
\end{itemize}
The first represents just the accuracy of a {\itshape random} classifier ({\bfseries Random}), whereas the second is typically referred to as {\bfseries NIR} -- No Information Rate or Null Information Rate. This second can also be considered as the result of the {\itshape a priori} classifier, a Bayesian classifier which assigns every object to the class $\omega_i$ which a priori probability $P(\omega_i)$ is maximum. $P(\omega_i)$ is estimated on the training set as:
\begin{equation}
P(\omega_i) = \frac{N_i}{N}
\end{equation}
where $N_i$ is the number of objects belonging to the $i$-th class, over the whole set of $N$ objects in the training set \cite{duda2006pattern}. While the intuition behind the first is trivial (just a random answer), for the second we exploit the a priori knowledge on the problem: if we know that one class is more frequent than another (for example the ``control'' class with respect to the ``disease'' class), then we can assign every testing object to such class. This represents another baseline, obtained with a very naive strategy which, at least, reflects the nature of the problem.

\section{Notation}
Before describing the test, let us introduce some notation. Given a dataset for a problem with $C$ classes, let us call $X^T$ and $X^E$ the training and the testing sets, respectively, composed by the objects ($x_i$) and the corresponding true labels ($y_i$):
$$
X^T = \{(x^T_1,y^T_1) \cdots (x^T_n,y^T_n)\}, \quad X^E = \{(x^E_1,y^E_1) \cdots (x^E_m,y^E_m)\}, \quad X^T \cap X^E = \emptyset
$$
Within the Cross Validation protocol, we use the training set $X^T$ to build the classifier $\mathcal{C}$; then we use the trained classifier to classify every object in the testing set $X^E$. The accuracy of the classifier $\mathcal{C}$ on $X^E$ can be estimated by counting the number of correctly classified testing objects:
\begin{equation}
t(\mathcal{C})=\sum_{i=1}^{m}\mathbf{I}(y^E_i=\mathcal{C}(x^E_i))
\end{equation}
where $\mathcal{C}(x^E_i)$ is the classification provided by the classifier $\mathcal{C}$ on the testing object $x^E_i$, and $\mathbf{I}(cond)$ is 1 if the condition ``$cond$'' is true, 0 otherwise. From $t$ we can compute the accuracy as:
\begin{equation}
acc(\mathcal{C})=\frac{t(\mathcal{C})}{m}
\end{equation}
The baselines against which we want to compare are represented by the accuracy of the {\bfseries random classifier} and the {\bfseries NIR value}. The former is easily defined as: 
\begin{equation}
accRand = \frac{1}{C}
\end{equation}
while for the latter we have:
\begin{equation}
NIR = \frac{\vert X^E_{a} \vert} {m}
\end{equation}
where $X^E_{a}$ is the subset of $X^E$ which contains only objects of class $a$, $a$ represents the most frequent class inside the training set $X^T$ and $\vert \; \vert$ represents the cardinality of a set.

\section{The test}
The test we describe here is also used in the R Caret Package \cite{Caret}, in particular to compare the observed accuracy to NIR. This package, and the test contained therein, has been widely applied in other works, such as \cite{Chekroud}. The test is based on a one-sided binomial test: even though there is a lack of studies specifically related to this aspect, the use of a binomial test can be easily supported by considering a correctly classified sample as a success. Indeed, it is not uncommon to model (or think of) a classification task as a binomial experiment \cite{tan,aronoff} (paragraph 4.6.1), in which we can have a {\itshape success} or a {\itshape failure} depending on a correct or uncorrect classification. In detail, a Binomial test \cite{Rosner} evaluates whether the observed proportion of successes is significantly different from the expected proportion, which in our case corresponds to the successes of the Random classifier accuracy or the NIR. Please note that this is not an inferential test and it directly returns a p-value. 

Given the classifier $\mathcal{C}$ and its corresponding accuracy on the testing set $acc(\mathcal{C})$, let us assume that $acc(\mathcal{C}) \geq NIR$ (or  $acc(\mathcal{C}) \geq accRand$ if our baseline is the random accuracy) and that we want to check if the improvement is due to chance or not. To this end we perform the \textbf{binomial one-tailed test} as follows:
\begin{itemize}
\item We set the baseline against which we want to compare: in particular we define the expected probability of success $p$ and the expected probability of failure $q$ as :
$$
p=NIR, \qquad q=1-p = 1-NIR
$$
(or $p=accRand, q=1-accRand$ for the test with respect to the random classifier).
\item We set the number of observed successes as $t(\mathcal{C})$, i.e. the number of objects in the testing set $X^E$ which have been correctly classified by $\mathcal{C}$.
\item We compute the p-value $pval$, which represents the probability that the increase in the accuracy of $\mathcal{C}$ with respect to the NIR (or the Random accuracy) is due to chance, as:
\begin{equation}
    pval=\sum_{k=t(\mathcal{C})}^{m}{\binom{m}{k}p^kq^{m-k}} 
   \end{equation}
This corresponds to a one-sided binomial test with parameters $m$, $p$ and $q$
\item If $pval\leq \alpha$ -- where usually $\alpha \leq 0.05$  -- then the classifier $\mathcal{C}$ is significantly better than the NIR classifier in terms of accuracy.
\end{itemize}

\section{Final Remarks}
\begin{itemize}
\item We can also compute the two-tailed p-value $pval2$, which is the probability that the difference between $acc(\mathcal{C})$ and the $NIR$ (or the Random accuracy) is due to chance (without assuming that $acc(\mathcal{C}) \geq NIR$ or  $acc(\mathcal{C}) \geq accRand$). In detail:
\begin{equation}
    pval2=2\sum_{k=t(\mathcal{C})}^{m}{\binom{m}{k}p^kq^{m-k}}.
\end{equation}
    \item Please note that in \cite{Rosner} they state that whenever $mpq\geq5$ the binomial distribution can be approximated to a normal one and a \textit{z-statistic can be computed}. (No mention of this has been made in \cite{Caret}). 
\item Matlab code is available from \url{https://profs.sci.univr.it/~bicego/code.html}.
\end{itemize}


\end{document}